\documentclass[aps,prd,10pt,twocolumn,groupedaddress,showpacs,showkeys]{revtex4-1}

\usepackage{amsmath}
\usepackage{amsfonts}
\usepackage{amssymb}
\usepackage{dsfont}
\usepackage{accents}
\usepackage[mathscr]{eucal}
\usepackage{braket}

\usepackage{graphicx}

\newcommand{\rmd}{\ensuremath{\mathrm{d}}}
\newcommand{\rme}{\ensuremath{\mathrm{e}}}
\newcommand{\rmi}{\ensuremath{\mathrm{i}}}

\newcommand{\Cb}{\ensuremath{{\tilde{C}}}}
\newcommand{\Vb}{\ensuremath{{\tilde{V}}}}
\newcommand{\Ub}{\ensuremath{{\tilde{U}}}}
\newcommand{\vb}{\ensuremath{{\bar{v}}}}
\newcommand{\ub}{\ensuremath{{\bar{u}}}}

\newcommand{\rs}{\ensuremath{r_{\rm s}}}
\newcommand{\rf}{\ensuremath{r_{\rm f}}}

\newcommand{\apr}{\ensuremath{a_{\rm{p}}}}

\usepackage[usenames,dvipsnames]{color}

\begin{document}

\title{A tensorial description of particle perception in black-hole physics}

\author{Luis C.\ Barbado}
\email[]{luis.cortes.barbado@univie.ac.at}
\affiliation{Quantenoptik, Quantennanophysik und Quanteninformation, Fakult\"at f\"ur Physik, Universit\"at Wien, Boltzmanngasse 5, 1090 Wien, Austria\\
Instituto de Astrof\'{\i}sica de Andaluc\'{\i}a (CSIC), Glorieta de la Astronom\'{\i}a, 18008 Granada, Spain}
\author{Carlos Barcel\'o}
\email[]{carlos@iaa.es}
\affiliation{Instituto de Astrof\'{\i}sica de Andaluc\'{\i}a (CSIC), Glorieta de la Astronom\'{\i}a, 18008 Granada, Spain}
\author{Luis J.\ Garay}
\email[]{luisj.garay@ucm.es}
\affiliation{Departamento de F\'{\i}sica Te\'orica II, Universidad Complutense de Madrid, 28040 Madrid, Spain\\
Instituto de Estructura de la Materia (CSIC), Serrano 121, 28006 Madrid, Spain}
\author{G.\ Jannes}
\email[]{gil.jannes@universidadeuropea.es}
\affiliation{Departamento de Ciencias y Tecnolog\'{\i}a, Universidad Europea de Madrid, Calle Tajo, 28670 Villaviciosa de Od\'on, Madrid, Spain}

\date{\today}

\begin{abstract}
In quantum field theory in curved backgrounds, one typically distinguishes between \emph{objective, tensorial,} quantities such as the Renormalized Stress-Energy Tensor (RSET) and \emph{subjective, non-tensorial,} quantities such as Bogoliubov coefficients which encode perception effects associated with the specific trajectory of a detector. In this work we propose a way to treat both objective and subjective notions on an equal tensorial footing. For that purpose, we define a new tensor which we will call the \emph{Perception Renormalized Stress-Energy Tensor} (PeRSET). The PeRSET is defined as the subtraction of the RSET corresponding to two different vacuum states. Based on this tensor we can define perceived energy densities and fluxes. The PeRSET helps to have a more organized and systematic understanding of various results in the literature regarding quantum field theory in black hole spacetimes. We illustrate the physics encoded in this tensor by working out various examples of special relevance.
\end{abstract}



\maketitle

\section{Introduction}

Most of the essential work on the detection of particles and energy in a curved spacetime dates from the 1970s.
It led to very important discoveries such as Parker's cosmological particle production~\cite{Parker1968,Parker1969}, Hawking's proposal that black holes should evaporate~\cite{Hawking1974}, and the Unruh effect for accelerated observers~\cite{Unruh1976}. Interest in these matters is continuously renewed though (see for example the recent works \cite{Barbado:2011dx, Barbado:2012pt, Smerlak:2013sga, Bianchi:2014qua, Singh:2014paa, Greenwood:2008zg, Anderson:2015iga, Singleton:2011vh, Crispino:2012zz, Singleton:2012zz, Good:2015jwa}).

Essentially, in the literature there exist two different approaches to describe a quantum radiation field in a curved spacetime: a) through the renormalized stress-energy tensor (RSET) and b) through Bogoliubov transformations and the perception effects associated with particle detectors. These two distinct approaches are treated quite separately in the literature. In this article, we will relate the two, providing a unified treatment of them. Our treatment rests on the introduction of a new (and generally defined) tensorial quantity, the \emph{Perception Renormalized Stress-Energy Tensor} (PeRSET). While there have been some previous attempts to deal with these two approaches simultaneously in concrete scenarios (see for example~\cite{Smerlak:2013sga}), this has still been done in a separate manner. In this work we present a new insightful way to look at these effects. 

The RSET is by definition a tensor, and can therefore be used to calculate physical, objective quantities such as energy densities and fluxes. These are obtained by contracting the RSET with appropriate four-velocity and surface-normal fields. Thus, they can be interpreted as the energy density and flux associated with a specific observer: an observer with a particular position and velocity. However, they do not depend on the observer's acceleration. For instance, the RSET of a field in a Minkowski spacetime and in the Minkowski vacuum state is zero, and since it is a tensor, it is zero independently of the acceleration of the observer. 

Therefore, the RSET does not describe the Unruh effect, the thermal bath perceived by an accelerated observer in the Minkowski vacuum~\cite{Unruh1976}. The Unruh effect can be accounted for by Bogoliubov transformations, and in principle be measured by e.g.\ an Unruh-DeWitt particle detector~\cite{Unruh1976,Dewitt:1979} (see also \cite{Louko:2007mu, Hodgkinson:2013tsa}). But these tools have the opposite drawback: Bogoliubov transformations and particle detector effects (and in particular those associated with accelerations) seem to lack the objective tensorial property of the RSET. This issue has been at the heart of an  ongoing controversy about whether Unruh radiation could really be detected at all, even in principle (see~\cite{Crispino:2007eb} for a review, and~\cite{Gelfer:2015voa} for a recent example), a view that we do not share, as it will become clear along this paper. So the RSET on the one hand, and perception effects through particle detectors on the other hand, encode different aspects of quantum field theory in curved backgrounds which, at first sight, seem to require separate treatment. 

A word of caution with respect to the terminology is in order here. For example the authors in~\cite{Smerlak:2013sga, Singh:2014paa} use the word ``perception'' for analyses based directly and only on the RSET. Perception is then understood as the detection of an objective quantity in the tensorial sense described earlier. Here, as in previous works~\cite{Barbado:2011dx, Barbado:2012pt, Barbado:2015zga}, we will reserve the word ``perception'' for those analyses in which, apart from the position and velocity of the observer, his acceleration also plays a role. This should not lead to any confusion about the reality of the quantities which we will describe. In fact, the idea that quantities associated with acceleration can also be described in a tensorial way is precisely our central objective in this manuscript.

We will construct a quantity, the Perception RSET (PeRSET), based on the calculation of the difference of the usual RSET in different vacua, and therefore maintaining its objective, tensorial property, but which at the same time will satisfactorily describe the full dependence of the perception on the state of motion of the observer, including his acceleration. The subtraction of the RSET in different vacua has been used before (see e.g.~\cite{Fabbri:2005mw,Firouzjaee:2015bqa}) but not in the form proposed here. We will find that, when defining the PeRSET, an already known quantity naturally appears: the effective temperature function introduced in~\cite{Barcelo:2010pj,Barcelo:2010xk}, which we will review in the next section. This function was introduced in the context of Bogoliubov transformations and particle perception. Its natural appearance in a quantity constructed from the RSET helps to unify both approaches.

The structure of this paper is the following. We will set the stage in Sec.~\ref{sec:prelim}, with some general preliminaries. In Sec.~\ref{sec:PeRSET}, we define the PeRSET, a tensorial quantity based on a comparison of the RSETs corresponding to different vacua, but nevertheless accounting for perception effects including the observer's acceleration. Sec.~\ref{sec:examples} contains several examples of the behavior of the PeRSET in different situations, and their physical interpretation. Finally, we summarize and discuss our results in Sec.~\ref{sec:summary}.

\section{Preliminaries}\label{sec:prelim}

We will consider a conformally invariant massless real scalar field in a $(1+1)$-dimensional spacetime. As an approximation to the $(3 + 1)$-dimensional case, this amounts to considering only the s-wave sector of the theory, and neglecting an effective potential term in the wave equation which is responsible for the backscattering of the field on the metric. This approximation has the important feature that it allows the analytic calculation of the RSET in a general vacuum state, something not possible in the $(3 + 1)$-dimensional case. Although the calculation of the RSET in the full $(3+1)$-dimensional will be more involved and only numerically computable, one would expect that this approximation captures many of the relevant features of the exact case.

Our field satisfies the massless Klein-Gordon equation, which in some fiduciary null coordinates $(\ub,\vb)$ reads
\begin{equation}
\square \phi=	\frac{\partial}{\partial \ub} \frac{\partial}{\partial \vb} \phi = 0.
\label{klein-gordon}
\end{equation}
Conformal invariance makes the~$\ub$ and~$\vb$ radiation sectors decouple. The general solution is then of the form \begin{equation}
\phi (\ub,\vb)= f(\ub) + g(\vb).
\label{general.solution}
\end{equation}

Also due to the conformal invariance of the $(1+1)$-Klein-Gordon equation, any relabeling $U=U(\ub )$, $V=V(\vb)$  can be associated with an expression of the $(1+1)$-metric written in these coordinates
\begin{equation}
\rmd s^2 = - C (U, V) \rmd U \rmd V,
\label{metric}
\end{equation} 
and a corresponding decomposition of the field in terms of the modes
\begin{equation}
\phi_{\omega }^U =\frac{1 }{ \sqrt{4\pi \omega }} \rme^{-\rmi \omega  U}, \quad
\phi_{\omega }^V =\frac{1 }{ \sqrt{4\pi \omega }} \rme^{-\rmi \omega  V},
\label{modes}
\end{equation}
which are orthonormal in the Klein-Gordon scalar product
\begin{equation}
\langle \phi_1, \phi_2 \rangle := - \rmi\left( - \int  \rmd U \phi_1  \stackrel{\leftrightarrow}{\partial}_U \phi_2^* + \int \rmd V \phi_1  \stackrel{\leftrightarrow}{\partial}_V \phi_2^* \right).
\label{scalar.product}
\end{equation}
After a canonical quantization procedure, we obtain natural annihilation $\hat a_{\omega }^U,\hat a_{\omega }^V$ and creation ${\hat a_{\omega }^U}^\dag,{\hat a_{\omega }^V}^\dag$  operators and a natural vacuum state $\ket{0}$  associated with these modes, which give rise to a Fock space and satisfy
\begin{equation}
\hat a_{\omega }^U |0   \rangle = 0, \quad \hat a_{\omega }^V |0  \rangle = 0.
\label{vacuum.definition}
\end{equation}

For example, in Minkowski spacetime, the selection of either Minkowski or Rindler null coordinates leads to Minkowski or Rindler vacuum states, respectively. The selection of either Eddington-Finkelstein or Kruskal-Szekeres coordinates for the outgoing~($U$) or the ingoing~($V$) null coordinates leads to the well-known Unruh, Boulware or Hartle-Hawking vacuum states~\cite{Fabbri:2005mw}. But infinitely many other vacuum states are possible, in particular also non-stationary vacuum states, such as the collapse vacuum state introduced in~\cite{Barbado:2011dx} (see also the discussion in~\cite{Anderson:2015iga}), or the pulsating vacuum state introduced in~\cite{Barbado:2011ai}.

Let us now review the two quantities that we will relate afterwards in Sec.\ref{sec:PeRSET}, namely the renormalized stress-energy tensor (RSET) and the effective temperature function.

\subsection{Renormalized stress-energy tensor}

For the approximate conformally-invariant theory we are considering, the components of the RSET in the vacuum state $\ket{0}$ acquire the well-known expressions \cite{Davies:1976ei,Davies:1977,Birrell:1982ix}
\begin{align}
\bra{0 } T_{U U} \ket{0 } & 
= \frac{1}{24 \pi C} \left[ \partial^2_U C - \frac{3}{2C} (\partial_U C)^2 \right], \label{rset_uu} \\
\bra{0 } T_{V V} \ket{0 } & 
= \frac{1}{24 \pi C} \left[ \partial^2_V C - \frac{3}{2C} (\partial_V C)^2 \right], \label{rset_vv} \\
\bra{0 } T_{U V} \ket{0 } & = \frac{1}{96 \pi} \partial_U \partial_V \log C.   
\label{rset_vu}
\end{align}

Given an observer with a trajectory~$(U(\tau),V(\tau))$, his four-velocity vector and normal (pointing in the direction of increasing $U$) are
\begin{equation}
 u^\mu=\left(\frac{\rmd U}{\rmd \tau},\frac{\rmd V}{\rmd \tau}\right),
 \quad
 n^\mu=\left(\frac{\rmd U}{\rmd \tau},-\frac{\rmd V}{\rmd \tau}\right).
 \label{vel-normal}
\end{equation}

The outgoing energy density and flux associated with this observer acquire the form
\begin{align}
\bra{0 } E\ket{0 }& :=  \bra{0 }{T}_{\mu \nu}\ket{0 } u^\mu u^\nu \nonumber \\
& =  \bra{0 }{T}_{U U}\ket{0 } \left(\frac{\rmd U}{\rmd \tau}\right)^2 +  \bra{0 }{T}_{V V}\ket{0 } \left(\frac{\rmd V}{\rmd \tau}\right)^2,\nonumber \\
\bra{0 } F\ket{0 } & :=  \bra{0 }{T}_{\mu \nu}\ket{0 } u^\mu n^\nu \nonumber \\
& = \bra{0 }{T}_{U U}\ket{0 }\left(\frac{\rmd U}{\rmd \tau}\right)^2  -  \bra{0 }{T}_{V V} \ket{0 }\left(\frac{\rmd V}{\rmd \tau}\right)^2 .
\end{align}
These quantities depend only on the position and the velocity of the observer. They do certainly not depend on the acceleration. In this sense, they fail to describe the perception of a detector, which depends strongly on its acceleration. As we already mentioned, in flat spacetime (in the Minkowski vacuum) the RSET vanishes identically and can therefore not account for the particle perception associated with an accelerated observer, i.e.\ the Unruh effect.

\subsection{Effective temperature function}

The particle perception of a generic observer can be described by means of non-tensorial quantities, such as the Bogoliubov coefficients between the modes defining the vacuum state and the modes to which the observer naturally couples. Based on the Bogoliubov transformations we can define the so-called effective temperature function or peeling function. This function was introduced in~\cite{Barcelo:2010pj,Barcelo:2010xk} and has been extensively used to analyze perception by different observers in various quantum vacua in~\cite{Barbado:2011dx,Barbado:2012pt}.

Given an observer with proper time~$\tau$ following a trajectory~$(U(\tau),V(\tau))$, with the quantum field in the vacuum state~$\ket{0}$, the effective temperature functions for the~$U$ and~$V$ radiation sectors are defined to be, respectively,
\begin{equation}
\kappa_U (\tau):= -\left. \frac{\rmd^2 U}{\rmd \tau^2} \middle/ \frac{\rmd U}{\rmd \tau} \right., \quad  \kappa_V (\tau):= -\left. \frac{\rmd^2 V}{\rmd \tau^2} \middle/ \frac{\rmd V}{\rmd \tau} \right.. 
\label{kappa}
\end{equation}
In~\cite{Barcelo:2010pj} it was proved that, when these functions remain constant for a sufficiently long period of time (controlled by an adiabaticity condition), the observer perceives a thermal spectrum of particles during this period (in the corresponding radiation sector) with temperature $T = |\kappa_U| / (2 \pi)$ and $T = |\kappa_V| / (2 \pi)$, respectively. For instance, in the case of an observer in the Minkowski vacuum state with uniform acceleration~$a$, these functions are simply $\kappa_U = -\kappa_V = a\ (= \rm{const})$, and thus the observer perceives a thermal bath with a temperature proportional to his acceleration in both sectors. Thus, the effective temperature function does indeed account for the Unruh effect.

\section{The perception~RSET (PeRSET)}\label{sec:PeRSET}

We define the PeRSET as a subtraction of the RSET in two vacuum states. In the perception question that we are dealing with, we have on the one hand the field in a concrete vacuum state, which we will denote by~$\ket{0}$. We will associate this first vacuum state with the (up to now generic) null coordinates~$(U, V)$. On the other hand, the observer interacts with the field following a specific trajectory. The second vacuum state will be the one locally perceived as vacuum for this particular observer. It will be associated with the null coordinates~$(\Ub, \Vb)$ and we will denote it by~$\ket{\tilde 0}$.

The definition of the PeRSET will then be the following:
\begin{equation}
\mathscr{T}_{\mu \nu} := \bra{0} T_{\mu \nu} \ket{0} - \bra{\tilde 0} T_{\mu \nu} \ket{\tilde 0}.
\label{perception_rset}
\end{equation}
Although we first need to characterize the vacuum state~$\ket{\tilde 0}$ and its corresponding RSET in order to make this definition precise, it is immediately obvious that $\mathscr{T}_{\mu \nu} = 0$ whenever $\ket{0}=\ket{\tilde 0}$, as required: any observer's perception is zero in his own local vacuum state.

In the following we will compare the usual RSET in the two different vacuum states, compute the subtraction, and find the relation between the PeRSET and the effective temperature function that we are looking for. But first, let us see how one can construct the local vacuum state for the observer whose perception we are considering.

\subsection{Construction of the local vacuum}\label{sec:Local_vacuum}
Now, how do we fix the null coordinates $(\Ub, \Vb)$ so that the
state~$\ket{\tilde 0}$ is really perceived as vacuum by our observer? As
we did in the previous sections, we select some initial fiduciary referential
coordinates $(\bar{u},\bar{v})$ to describe the specific
effectively $(1+1)$-dimensional spacetime under consideration. Let
$\bar{v}=f(\bar{u})$ be the time-like trajectory of the local observer.
This observer can use his proper time to label the events determined by
him crossing the different $\bar{u}$ and $\bar{v}$~rays. In other words,
his trajectory can be parametrized in terms of his proper time:
$\bar{u}=g_{\bar{u}}(\tau-\tau_0)$, $\bar{v}=g_{\bar{v}}(\tau-\tau_0)$.
Then, he can use these very functions as defining the observer local
vacuum through the selection of coordinates $( \tilde{U}=
\tilde{U}(\bar{u}),\tilde{V}= \tilde{V}(\bar{v}) )$  such that
$\bar{u}=g_{\bar{u}}(\tilde{U})$, $\bar{v}=g_{\bar{v}}(\tilde{V})$.
The functions $g_{\bar{u}}$ and $g_{\bar{v}}$ can be found by realizing
that, along the trajectory:
\begin{eqnarray}
\rmd\tau^2= -C(\bar{u},\bar{v})\rmd\bar{u}\rmd\bar{v}= -C(\bar{u},f(\bar{u}))
\frac{ \rmd f(\bar{u}) }{ \rmd\bar{u}} \rmd\bar{u}^2,
\end{eqnarray}
and equivalently for $\bar{v}$. In this way we obtain
\begin{eqnarray}
&&\rmd \tilde{U} = \left[ C(\bar{u},f(\bar{u}))\frac{ \rmd f(\bar{u}) }{
\rmd\bar{u}} \right]^{1/2}\rmd\bar{u},
\\
&&\rmd \tilde{V} = \left[ C(f^{-1}(\bar{v}),\bar{v})\frac{ \rmd
f^{-1}(\bar{v}) }{ \rmd\bar{v}} \right]^{1/2}\rmd\bar{v},
\end{eqnarray}
which can be integrated. It is worth noting that, at
any regular point on the trajectory of the observer, $(
\bar{u}_0,\bar{v}_0 )$ or  equivalently $( \tilde{U}_0,\tilde{V}_0 )$, the
metric will have a locally
Minkowskian form $\rmd s^2=-\rmd\tilde{U} \rmd\tilde{V}$. Note that
the factor $\displaystyle{(\rmd f^{-1}/\rmd\bar{v})\big|_0}$ is precisely
equal to the inverse of
$\displaystyle{(\rmd f/\rmd\bar{u})\big|_{0}}$. These two factors
correspond to left-going and right-going Doppler factors, respectively (a
local change of velocity does not change the local form of the metric).

Of course, in general, outside the trajectory the metric will not have a
Minkowskian form. Note also that the selected null coordinates are not
null normal coordinates, that is, they are not adapted to the local
free-fall, but to the trajectory of the observer, which in general is not
geodesic. It is also interesting to realize that in order to calculate the
PeRSET in a point $( \bar{u}_0,\bar{v}_0 )$ one would only need $\bar{v}=
f(\bar{u})$ in that point up to its third derivative [look e.g. at
expressions~(\ref{rel_rset_uu})--(\ref{rel_rset_uv}) below]. For instance, in trying to generalize this local-vacuum construction to the general $(3+1)$-dimensional case, one should do it point by point along the trajectory. Backscattering would cause that the local-vacuum modes associated to one point of the trajectory would be different from those associated to other points. This will add to the difficulty of calculating a RSET in the general case.

\subsection{Comparing the RSET in different vacua}\label{sec:comparing-RSETs}

Let us consider the null coordinate system~$(\Ub,\Vb)$. In this coordinate system the metric reads
\begin{equation}
\rmd s^2 = - \Cb (\Ub, \Vb) \rmd \Ub \rmd \Vb.
\label{new_metric}
\end{equation}
By comparing with~(\ref{metric}), one can trivially see that
\begin{equation}
C = \Cb \frac{\rmd \Ub}{\rmd U}\frac{\rmd \Vb}{\rmd V}. 
\label{C_comparison}
\end{equation}
As in the previous section, we can perform a Fock quantization in terms of the natural modes associated with these new coordinates. This leads to the vacuum state $\ket{\tilde 0}$. We want to relate the components of the RSET in~(\ref{rset_uu})--(\ref{rset_vu}) with the components that one can compute in an analogous way in the new coordinate system~$(\Ub,\Vb)$ and the new associated vacuum~$\ket{\tilde 0}$. From the relation between the conformal factors~(\ref{C_comparison}) we can compute the $U$-derivatives
\begin{align}
\partial_U C = & \left( \partial_\Ub \Cb + \Cb \kappa^\Ub_U \right) \left( \frac{\rmd \Ub}{\rmd U} \right)^2 \frac{\rmd \Vb}{\rmd V}, \\
\frac{1}{C} \partial^2_U C = & \Bigg[ \frac{1}{\Cb} \partial^2_\Ub \Cb + \frac{3}{\Cb} (\partial_\Ub \Cb) \kappa^\Ub_U + \frac{\rmd \kappa^\Ub_U}{\rmd \Ub} \nonumber \\
& + 2 (\kappa^\Ub_U)^2 \Bigg] \left( \frac{\rmd \Ub}{\rmd U} \right)^2, \\
\frac{1}{C^2} (\partial_U C)^2 = & \Bigg[ \frac{1}{\Cb^2} (\partial_\Ub \Cb)^2 + (\kappa^\Ub_U)^2 \nonumber \\
& + \frac{2}{\Cb} (\partial_\Ub \Cb) \kappa^\Ub_U \Bigg] \left( \frac{\rmd \Ub}{\rmd U} \right)^2,
\end{align}
where we have defined the ``relative effective temperature function'' between the two vacua [in a way analogous to the effective temperature function~$\kappa_U$ itself, see eq.~(\ref{kappa})] as
\begin{equation}
\kappa^\Ub_U := -\left. \frac{\rmd^2 U}{\rmd \Ub^2} \middle/ \frac{\rmd U}{\rmd \Ub} \right. = \left. \frac{\rmd^2 \Ub}{\rmd U^2} \middle/ \left( \frac{\rmd \Ub}{\rmd U} \right)^2 \right..
\label{kappa_definition}
\end{equation}
Identical expressions for the $V$-derivatives and the function~$\kappa^\Vb_V$ also hold true after replacing~$U \leftrightarrow V$ and~$\Ub \leftrightarrow \Vb$. Substituting these expressions in~(\ref{rset_uu})--(\ref{rset_vu}), we find that
\begin{multline}
\bra{0 } T_{U U} \ket{0 } = \Bigg[ \bra{\tilde 0 } T_{\Ub \Ub} \ket{\tilde 0 } \\
+ \frac{1}{24 \pi} \left( \frac{1}{2} (\kappa^{\Ub}_U)^2 + \frac{\rmd \kappa^\Ub_U}{\rmd \Ub} \right) \Bigg] \left( \frac{\rmd \Ub}{\rmd U} \right)^2,
\label{rset_comp_Uu}
\end{multline}
\begin{multline}
\bra{0 } T_{V V} \ket{0 } = \Bigg[ \bra{\tilde 0} T_{\Vb \Vb} \ket{\tilde 0} \\
+ \frac{1}{24 \pi} \left( \frac{1}{2} (\kappa^{\Vb}_V)^2 + \frac{\rmd \kappa^\Vb_V}{\rmd \Vb} \right) \Bigg] \left( \frac{\rmd \Vb}{\rmd V} \right)^2,
\label{rset_comp_vv}
\end{multline}
\begin{equation}
\bra{0 } T_{U V} \ket{0 } = \bra{\tilde 0} T_{\Ub \Vb} \ket{\tilde 0} \frac{\rmd \Ub}{\rmd U} \frac{\rmd \Vb}{\rmd V},
\label{rset_comp_vu}
\end{equation}
where the last equation comes from
\begin{align}
\partial_U \partial_V \log C & = \partial_U \partial_V \left( \log \Cb + \log \frac{\rmd \Ub}{\rmd U} + \log \frac{\rmd \Vb}{\rmd V} \right) \nonumber \\
& = (\partial_\Ub \partial_\Vb \log \Cb) \frac{\rmd \Ub}{\rmd U} \frac{\rmd \Vb}{\rmd V}.
\end{align}
Note that, due to the tensorial transformation of the RSET, we have
\begin{align}
\bra{\tilde 0} T_{\Ub \Ub} \ket{\tilde 0} \left( \frac{\rmd \Ub}{\rmd U} \right)^2 = \bra{\tilde 0} T_{U U} \ket{\tilde 0}, \label{rset_transform_uu} \\
\bra{\tilde 0} T_{\Vb \Vb} \ket{\tilde 0} \left( \frac{\rmd \Vb}{\rmd V} \right)^2 = \bra{\tilde 0} T_{V V} \ket{\tilde 0}, \label{rset_transform_vv} \\
\bra{\tilde 0} T_{\Ub \Vb} \ket{\tilde 0} \frac{\rmd \Ub}{\rmd U} \frac{\rmd \Vb}{\rmd V} = \bra{\tilde 0} T_{U V} \ket{\tilde 0}. \label{rset_transform_vu}
\end{align} 
Therefore, we finally obtain
\begin{align}
\bra{0} T_{U U} \ket{0} = & \bra{\tilde{0}} T_{U U} \ket{\tilde{0}} \nonumber \\
& + \frac{1}{24 \pi} \left[ \frac{1}{2} (\kappa^{\Ub}_U)^2 + \frac{\rmd \kappa^\Ub_U}{\rmd \Ub} \right] \left( \frac{\rmd \Ub}{\rmd U} \right)^2, \label{rel_rset_uu}
\\
\bra{0} T_{V V} \ket{0} = & \bra{\tilde{0}} T_{V V} \ket{\tilde{0}} \nonumber \\
& + \frac{1}{24 \pi} \left[ \frac{1}{2} (\kappa^{\Vb}_V)^2 + \frac{\rmd \kappa^\Vb_V}{\rmd \Vb} \right] \left( \frac{\rmd \Vb}{\rmd V} \right)^2, \label{rel_rset_vv}
\\
\bra{0} T_{U V} \ket{0} = & \bra{\tilde{0}} T_{U V} \ket{\tilde{0}}. \label{rel_rset_uv}
\end{align}
Note that the difference between the RSETs associated with different vacua depends on the relative functions $U=U(\tilde{U})$, $V=V(\tilde{V})$ up to their third derivatives. Similar formal expressions can be found in the analysis of state purification done in~\cite{Bianchi:2014qua}.

\subsection{PeRSET and perceived energy density and flux}

Now that we have compared the~RSET in the two different vacua, with the expressions~(\ref{rel_rset_uu})--(\ref{rel_rset_uv}) we can compute the components of the PeRSET from the definition~\eqref{perception_rset}:
\begin{align}
\mathscr{T}_{U U} & = \bra{0} T_{U U} \ket{0} - \bra{\tilde 0} T_{U U} \ket{\tilde 0} \nonumber \\
& = \frac{1}{24 \pi} \left( \frac{1}{2} \kappa^2_U + \frac{\rmd \kappa_U}{\rmd \tau} \right) \left( \frac{\rmd U}{\rmd \tau} \right)^{-2}, \label{perc_rset_uu} \\
\mathscr{T}_{V V} & = \bra{0} T_{V V} \ket{0} - \bra{\tilde 0} T_{V V} \ket{\tilde 0} \nonumber \\
& = \frac{1}{24 \pi} \left( \frac{1}{2} \kappa^2_V + \frac{\rmd \kappa_V}{\rmd \tau} \right) \left( \frac{\rmd V}{\rmd \tau} \right)^{-2}, \label{perc_rset_vv} \\
\mathscr{T}_{U V} & = \bra{0} T_{U V} \ket{0} - \bra{\tilde 0} T_{U V} \ket{\tilde 0} = 0 \label{perc_rset_vu}.
\end{align}
In these expressions, the effective temperature function $\kappa_U$ appears. This is because the vacuum $\ket{\tilde 0}$ is associated with coordinates such that $\rmd\tilde{U}=\rmd\tau$, as we have seen above, and therefore the quantity $\kappa^\Ub_U$ in~(\ref{kappa_definition}) is equal to the effective temperature function~$\kappa_U$ in~(\ref{kappa}) for our observer. The same comment applies to $\kappa^\Vb_V = \kappa_V$.

Finally, we will compute the perceived energy density and flux for the observer, whose four-velocity and normal are given in~(\ref{vel-normal}). These quantities yield
\begin{align}
\mathscr{E} & := \mathscr{T}_{\mu \nu} u^\mu u^\nu = \mathscr{T}_{U U} u^U u^U + \mathscr{T}_{V V} u^V u^V\nonumber \\
& = \frac{1}{24 \pi} \left[ \frac{1}{2} ( \kappa^2_U + \kappa^2_V) + \frac{\rmd \kappa_U}{\rmd \tau} + \frac{\rmd \kappa_V}{\rmd \tau} \right], \label{perceived_E} \\
\mathscr{F} & := \mathscr{T}_{\mu \nu} u^\mu n^\nu = \mathscr{T}_{U U} u^U n^U + \mathscr{T}_{V V} u^V n^V\nonumber \\
& = \frac{1}{24 \pi} \left[ \frac{1}{2} ( \kappa^2_U - \kappa^2_V) + \frac{\rmd \kappa_U}{\rmd \tau} - \frac{\rmd \kappa_V}{\rmd \tau} \right]. \label{perceived_F}
\end{align}
These expressions constitute our final result relating the PeRSET (obtained from two~RSETs) and the effective temperature function (first introduced in the framework of Bogoliubov transformations). Formally similar expressions can be found in \cite{Fabbri:2005mw,Firouzjaee:2015bqa}, but coming from a different construction and with a different physical motivation. Although these expressions already recall well-known physical laws, their meaning will be much clearer when we proceed with some examples of vacuum states and observer trajectories in Sec.~\ref{sec:examples}. But to emphasize the crucial point so far: we have proved that the PeRSET, apart from its obvious meaning as the difference between the RSET in two different vacuum states, can also account for perception aspects which involve up to the third derivative of the observer trajectory.

\section{Examples}\label{sec:examples}

In this section we consider a few concrete examples of geometries and observers of special interest, compute the PeRSET for them, and discuss the physical interpretation of the results.

This physical interpretation relies on a conceptualization and separation of the Unruh and Hawking effects that the present authors have argued for in~\cite{Barbado2016}. Thus, before entering into the details of the examples, we shall provide a brief summary of this interpretation.

\subsection{Hawking vs Unruh effects}\label{H-vs-U}

The conformal invariance of the effective description that we consider allows to encode all of the perception properties in two effective temperature functions, one for the ingoing sector and another for the outgoing sector. In~\cite{Barbado:2011dx} it was proved, for the outgoing effective temperature function, that it can be written analytically as a series of terms and factors with direct physical interpretations ---the ingoing sector has an equivalent formula which is given in~\cite{Barbado2016}---. In~\cite{Barbado2016} we further argue that the presence of an asymptotic region permits to separate the total effective temperature function into two different contributions, which can be associated with the Unruh and the Hawking effect, respectively. We argue that a separation free of inconsistencies should necessarily associate the Unruh effect with the acceleration of the observer \emph{with respect to the asymptotic region}, and not with respect to the local free-fall reference frame as is commonly assumed. If one is only interested in calculating the perception of a specific observer in a specific vacuum state, then this interpretation is equivalent to the more standard one. However, if one wants to go further and consider the backreaction of the radiation field on the very trajectory of the observer, then these two interpretations lead to different predictions. 

A physically relevant situation can help to clarify the issue. Imagine a detector set up at a fixed radial position very close to the horizon of an evaporating black hole, that is, in the Unruh state. The detector perceives a thermal emission from the black hole with an enormous temperature: Hawking's temperature multiplied by the very large blue-shift factor associated with the radial position just outside the horizon. But how does the radiation field act on the detector trajectory? The standard interpretation says that this detector's perception is caused by the Unruh effect: The Unruh vacuum is almost vacuum for a free-falling observer at the horizon; the detector in the fixed radial position is therefore strongly accelerating with respect to the free-fall frame; thus, the detector experiences a large Unruh effect. The back-reaction associated with this Unruh effect must introduce an additional force which tries to diminish the acceleration of the detector, and thereby pushes the detector towards the horizon. In this standard interpretation, the radiation field thus creates an additional force towards the black hole, which should be compensated for the detector to remain static outside the horizon. We argue that this tendency is in contradiction with the outgoing flux of particles detected by the observer.

Our interpretation, on the contrary, implies that the observer is experiencing only a Hawking effect. There is no Unruh effect since the detector cannot modify the structure of the field at infinity given that it is at rest with respect to the asymptotic region. Thus, the action of the radiation field on the detector would have a distinct buoyant effect and in fact would push the detector away from the horizon. In other words, the radiation field actually helps to maintain the detector in a static position just outside the black hole horizon. In the following we will stick to this interpretation (see~\cite{Barbado2016} for further details).

\subsection{Unruh effect in Minkowski spacetime}

Consider the $(1+1)$-Minkowski spacetime described in some inertial coordinates~$(t,x)$. The metric reads:
\begin{equation}
\rmd s^2 = -\rmd t^2 + \rmd x^2.
\label{minkowski}
\end{equation}

We fix the state of the field to be the Minkowski vacuum state, which is the vacuum state associated with the null coordinates $U := t - x$ and $V := t + x$. For a given observer following a trajectory~$(t(\tau), x(\tau))$, using the metric~(\ref{minkowski}) and the definitions~(\ref{kappa}), it is easy to see that
\begin{equation}
\kappa_U = - \kappa_V = \apr = \frac{\ddot x}{\sqrt{1+{\dot x}^2}},
\label{kappa_unruh}
\end{equation}
where the dot denotes derivation with respect to proper time, and~$\apr$ is the proper acceleration of the observer. Using this result in the expressions for the perceived energy~(\ref{perceived_E}) and flux~(\ref{perceived_F}), we get
\begin{equation}
\mathscr{E} = \frac{\apr^2}{24 \pi}, \quad \quad \mathscr{F} = \frac{{\dot a}_{\rm p}}{12 \pi}.
\label{perception_unruh}
\end{equation}

We can see that we obtain the result expected from the Unruh effect: the total energy perceived is given by $\mathscr{E} = (\pi/6) T_{\rm U}^2$, with $T_{\rm U} := |\apr|/(2 \pi)$ the Unruh temperature. This is the Stefan-Boltzmann law for $1+1$~dimensions. We also obtain a flux proportional to and in the same direction as the jerk (i.e., the time derivative of the acceleration), in remarkable agreement with the formal expression of the Abraham-Lorentz reaction force of classical electrodynamics (see e.g.~\cite{Jackson}). It is worth stressing the clear separation that we find: The energy density perceived is fully determined by the acceleration, while the flux perceived is fully determined by the jerk.

\subsection{Black hole spacetimes and Hawking radiation}

Let us now consider the Schwarzschild spacetime outside a spherically symmetric black hole of mass~$M$. We will perform all the calculations with the Schwarzschild metric, but they can easily be generalized to any spherically symmetric metric representing a black hole.

The exterior Schwarzschild metric is
\begin{align}
\rmd s^2 & = - \left(1 - \frac{2 M}{r} \right) \rmd t^2 + \frac{1}{1 - \frac{2M}{r}} \rmd r^2 \nonumber \\
& = \left(1 - \frac{2 M}{r} \right) [-\rmd t^2 + (\rmd r^*)^2],
\label{schwarzschild}
\end{align}
where~$r^* := r + 2M \log[r/(2M) - 1]$ is the tortoise coordinate.

In this spacetime, we will consider three different families of observers:

\begin{itemize}
	\item
		Static observers at a radius~$\rs$. Their trajectories are given by:
		\begin{equation}
			r = \rs, \quad \quad t = \frac{\tau}{\sqrt{1- \frac{2M}{\rs}}}.
		\label{static}
		\end{equation}
		
	\item
		Free-falling observers left to fall (with zero initial velocity) from a radius~$r_0$. Their trajectories are obtained by integrating the geodesic equation corresponding to the metric~(\ref{schwarzschild}). This yields the first order derivatives
		\begin{equation}
		\frac{\rmd r}{\rmd \tau} = - \sqrt{\frac{2M}{r} - \frac{2M}{r_0}}, \quad \frac{\rmd t}{\rmd \tau} = \frac{\sqrt{1 - \frac{2M}{r_0}}}{1 - \frac{2M}{r}}.
		\label{free-falling}
		\end{equation}
		The integration of these equations cannot be written explicitly as~$(t(\tau), r(\tau))$, but the first derivatives are sufficient to compute the values of~$\kappa_{V}$ and~$\kappa_{U}$ (see~\cite{Barbado:2011dx} for more details).

	\item
		Observers following the so-called \emph{quantum-frictionless} trajectories, introduced in~\cite{Barbado:2015zga}. Observers moving along these trajectories do not perceive Unruh effect either in the outgoing or in the ingoing radiation sector. This means that, in the corresponding sector, the observers only perceive the radiation coming from external sources (the stellar object or the asymptotic region), modified by a blueshift factor due to their position and a Doppler factor due to their velocity.
		
\end{itemize}

In the following, we will discuss the perceived energy and flux for the first two families of trajectories in the Boulware and Unruh vacuum states. For convenience in the presentation, we will treat the quantum frictionless trajectories in a different subsection.

\subsubsection{Boulware vacuum state}

In the Boulware vacuum state there is no emission of radiation by the black hole, and there is no radiation coming from the asymptotic region either. This vacuum state is associated with the Eddington-Finkelstein null coordinates:
\begin{equation}
	U = t - r^*, \quad V = t + r^*.
\label{eddington-finkelstein}
\end{equation}

\paragraph{Static observers}

For static observers, it is easy to see that~$\mathscr{E} = \mathscr{F} = 0$: Since there are no sources of radiation, and since they have no acceleration with respect to infinity (no Unruh effect), static observers in the Boulware vacuum do not perceive any radiation at all. Of course, static observers only exist, strictly, outside the horizon.

\paragraph{Free-falling observers}

Free-falling observers, on the contrary, will perceive radiation in the Boulware vacuum state. Since there are no sources of radiation, this perception must be purely due to the Unruh effect. Using the equations~(\ref{free-falling}) we obtain the following results for the perceived energy density and flux:
\begin{equation}
\mathscr{E} = \frac{M}{48 \pi r_0} \frac{\big(4r/M - 9 r_0 / (2M) - 6\big) r + 7 r_0}{\big(r/(2M) - 1 \big)^2 r^3},
\label{energy_boulware_falling}
\end{equation}
\begin{align}
\mathscr{F} =\ & \frac{M}{24 \pi r_0} \sqrt{\frac{r_0}{2M} - 1}\sqrt{\frac{r_0}{r} - 1} \nonumber \\
& \times \frac{3-5r/M+11\big(r/(2M)\big)^2-4\big(r/(2M)\big)^3}{\big(r/(2M)-1\big)^4r^2}.
\label{flux_boulware_falling}
\end{align}
It is easy to see that both quantities are always negative and diverge at the horizon crossing. This divergence appears from the fact that, when approaching the horizon, the acceleration of the observer with respect to the asymptotic region, which determines the Unruh effect, diverges for the outgoing radiation sector. At the initial position $r=r_0$, the flux vanishes while the energy density takes the value~$\mathscr{E} =  M/\left[2r_0^3(2M-r_0)\right]$.

\begin{figure}[h]
\centering
\includegraphics[width=8cm]{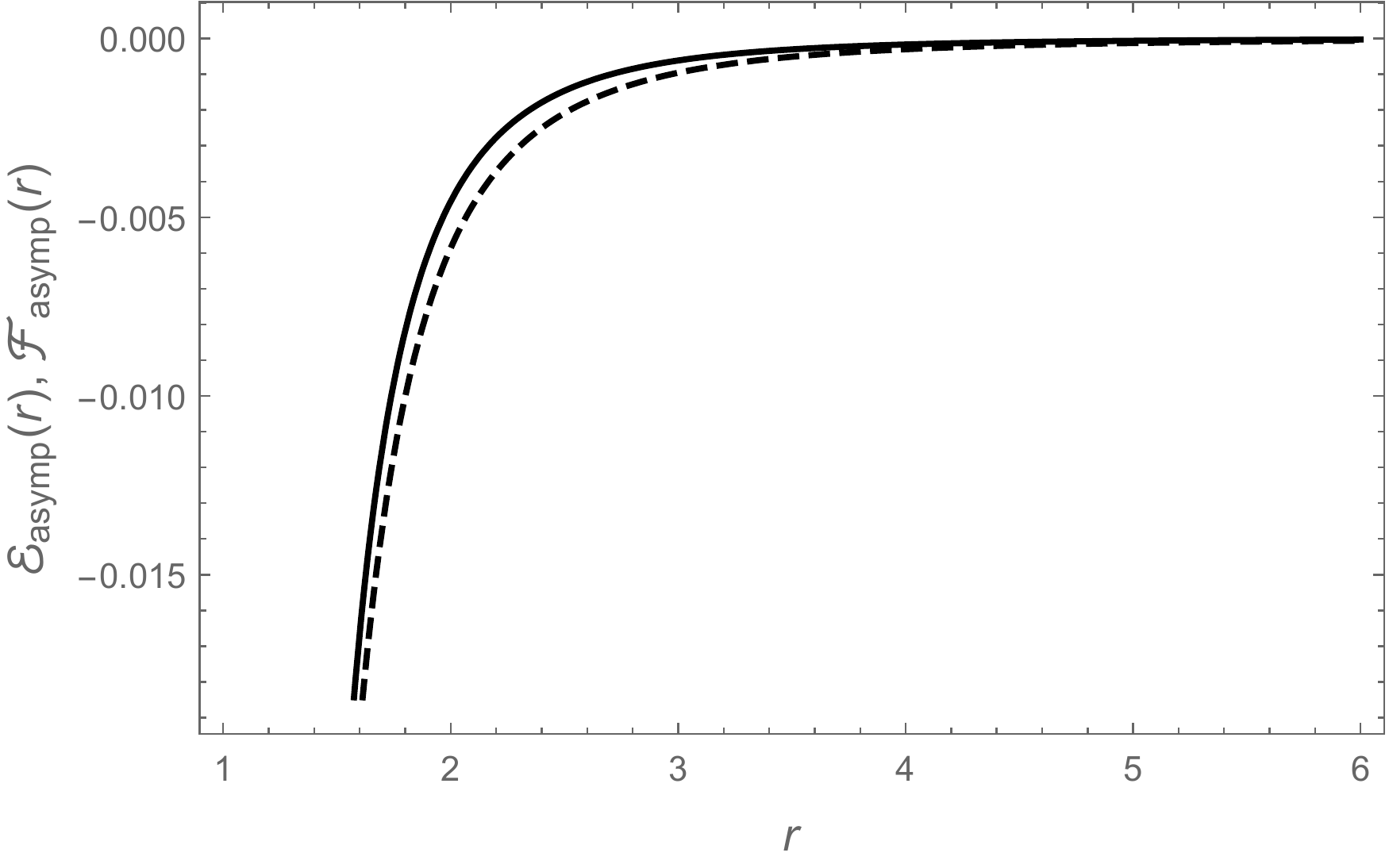}
\caption{Perceived energy density $\mathscr{E}_{\rm asymp}$ (solid line) and flux~$\mathscr{F}_{\rm asymp}$ (dashed line) as a function of~$r$ for free-falling observers from infinity in the Boulware vacuum state. We use $2M=1$~units.}
\label{fig_boulware_falling}
\end{figure}

A particularly interesting case is when~$r_0 \to \infty$, that is, the observer starts falling from the asymptotic region. In this case, the expressions simplify to:
\begin{equation}
\mathscr{E}_{\rm asymp} = \frac{M}{48 \pi} \frac{7 - 9r/(2M)}{\big(r/(2M) - 1 \big)^2 r^3},
\label{energy_boulware_falling_asymp}
\end{equation}
\begin{equation}
\mathscr{F}_{\rm asymp} = \frac{1}{48 \pi} \frac{3 - 2r/M}{r^2 \big( r/(2M) - 1 \big)^2 \sqrt{r / (2M)}}.
\label{flux_boulware_falling_asymp}
\end{equation}
In Fig.~\ref{fig_boulware_falling} we plot the values of~$\mathscr{E}_{\rm asymp}$ and~$\mathscr{F}_{\rm asymp}$ as a function of~$r$.

\subsubsection{Unruh vacuum state} \label{subsec:unruh}

In the Unruh vacuum state the black hole emits Hawking radiation to the asymptotic region, with the Hawking temperature~$T_{\rm H} := 1/(8 \pi M)$. This vacuum state is associated with the null coordinates:
\begin{equation}
	U = -4 M \rme^{-(t-r^*)/(4M)}, \quad V = t+r^*.
\label{unruh_ef_coordinates}
\end{equation}

\paragraph{Static observers}

Static observers in this case just perceive the flux of radiation emitted by the black hole, corrected with the blueshift factor corresponding to their position:
\begin{equation}
\mathscr{E} = \mathscr{F} = \frac{1}{768 \pi M^2 ( 1 - 2M/\rs )} = \frac{\pi}{12 ( 1 - 2M/\rs )} T_{\rm H}^2.
\label{unruh_static}
\end{equation}
Within our interpretation, we find that static observers at any radius only perceive the radiation present due to external sources, nothing more; that is, they perceive the same radiation perceived by static observers in the asymptotic region (who clearly can perceive nothing but the radiation due to external sources~\cite{Barbado:2015zga, Barbado2016}), just adequately blue-shifted according to their position. One can easily check that this is always the case, not only for Boulware or Unruh vacuum states, but regardless of the state considered. Indeed, for static observers in the asymptotic region the proper time is $\rmd \tau_{\rm asymp} = \rmd t$, while for static observers at any finite radial position~$\rs$ is $\rmd \tau_{\rm s} = \sqrt{1 - 2M/\rs}\ \rmd t$. From the expressions of the perceived energy density~(\ref{perceived_E}) and flux~(\ref{perceived_F}), and of the effective temperature function~(\ref{kappa}), it is then clear that (with obvious notation) $\mathscr{E}_{\rm s} = \mathscr{E}_{\rm asymp} / ( 1 - 2M/\rs )$ and $\mathscr{F}_{\rm s} = \mathscr{F}_{\rm asymp} / ( 1 - 2M/\rs )$. This illustrates the interpretation mentioned in Sec.~\ref{H-vs-U} above that the Unruh effect depends on the acceleration with respect to the asymptotic region, and is therefore necessarily absent for static observers.

\paragraph{Free-falling observers}

In the case of free-falling observers, the general expressions for the perceived energy density and flux, in terms of the starting radial position~$r_0$ and the radius~$r$, are very complicated for the Unruh vacuum state. However, there are three physically relevant limits in which the expressions hugely simplify.

\begin{figure}[h]
\centering
\includegraphics[width=8cm]{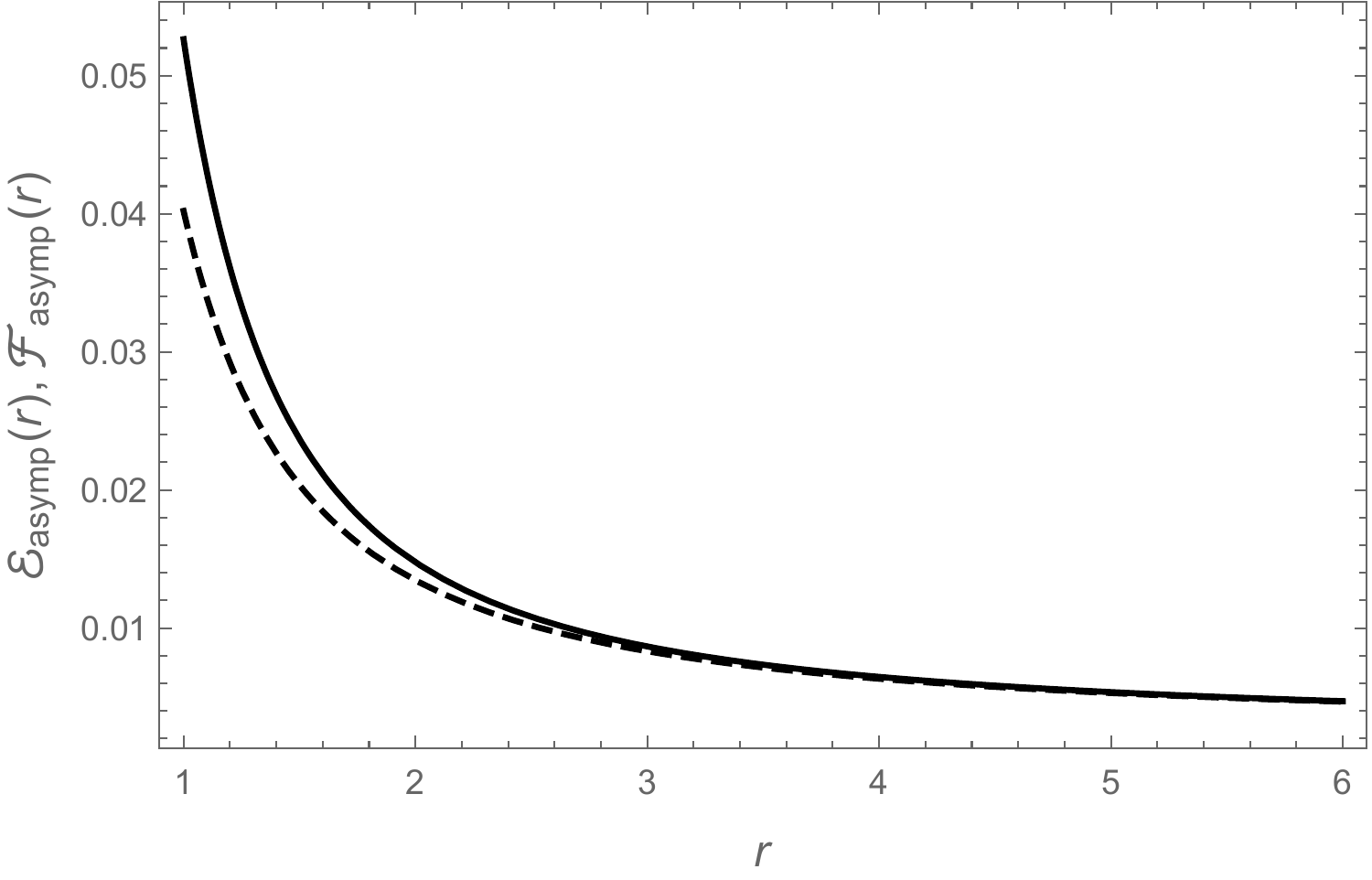}
\caption{Perceived energy density $\mathscr{E}_{\rm asymp}$ (solid line) and flux~$\mathscr{F}_{\rm asymp}$ (dashed line) as a function of~$r$ for free-falling observers from infinity in Unruh vacuum state. At horizon crossing they reach finite values of the same order of magnitude that the Hawking energy and flux detected at infinity. We use $2M=1$~units.}
\label{fig_unruh_falling}
\end{figure}

The first is the limit~$r_0 \to \infty$ (observer free-falling from the asymptotic region). In that limit, the expressions are:
\begin{align}
\mathscr{E}_{\rm asymp} =\ & \frac{1}{768 \pi M^2} \left[14 \left(\frac{2M}{r}\right)^3 + 10 \left(\frac{2M}{r}\right)^2 + \frac{12M}{r} \right. \nonumber \\
& \left. + \frac{2 + 4 \sqrt{r/(2M)}+r/(2M)}{\left(1 + \sqrt{r/(2M)} \right)^2} \right],
\label{energy_unruh_falling_asymp}
\end{align}
\begin{multline}
\mathscr{F}_{\rm asymp} = \frac{1}{192 \pi r^2 \sqrt{r/(2M)}} \\
\times \left[12 + \frac{r}{2M} \left( 8 + \frac{2r}{M} + \frac{(r/(2M))^{5/2}}{\left(1 + \sqrt{r/(2M)} \right)^2} \right) \right].
\label{flux_unruh_falling_asymp}
\end{multline}
We plot these quantities in~Fig.~\ref{fig_unruh_falling}. One can easily check that in the limit~$r \to \infty$ the quantities tend to~$\mathscr{E}_{\rm asymp} = \mathscr{F}_{\rm asymp} = 1/(768 \pi M^2) = (\pi / 12) T_{\rm H}^2$, which reproduces the result of the Hawking radiation flux. 
However, as the trajectory approaches the black hole the quantities start to increase, until reaching a finite limit when crossing the horizon, given by~$\mathscr{E}_{\rm asymp} \to 127/(3072 \pi M^2)$ and~$\mathscr{F}_{\rm asymp} \to 97/(3072 \pi M^2)$. 
These finite results agree with the finite results for the effective temperature functions that we found in~\cite{Barbado2016} for the same limit. This slight increase in energy and flux is consistent with the results found in~\cite{Barbado:2011dx} and later confirmed in~\cite{Smerlak:2013sga} showing that the value of $\kappa_U$ increases from the Hawking value at infinity to four times this value at the horizon crossing. Here the precise factor of 4 found in those works is not directly seen because the perceived energies and fluxes incorporate not only $\kappa_U$ but also $\kappa_V$ and their respective derivatives.

\begin{figure}[h]
\centering
\includegraphics[width=8cm]{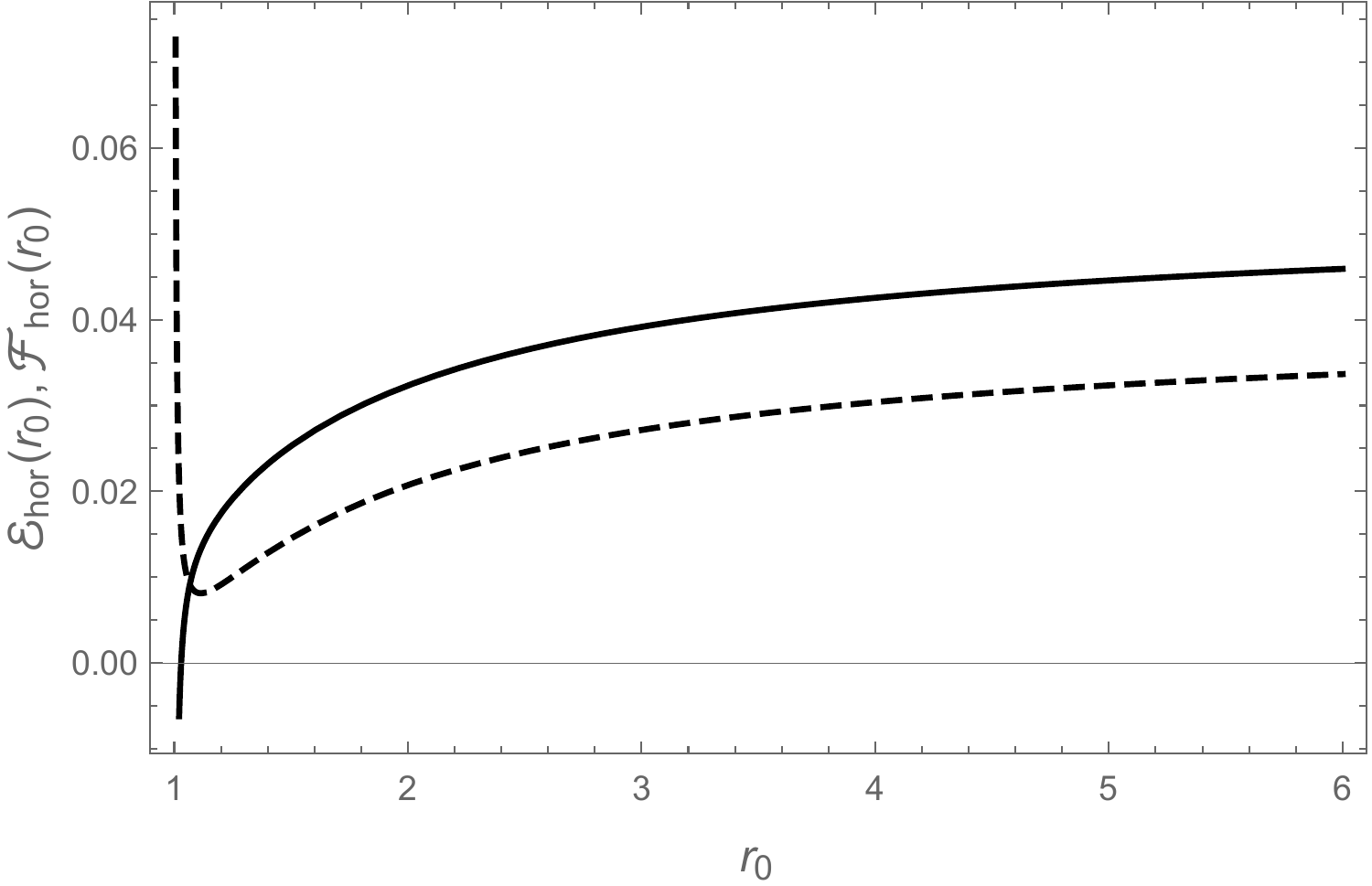}
\caption{Perceived energy density $\mathscr{E}_{\rm hor}$ (solid line) and flux~$\mathscr{F}_{\rm hor}$ (dashed line) as a function of the releasing radius~$r_0$ for free-falling observers in Unruh vacuum state. We use $2M=1$~units.}
\label{fig_unruh_hor}
\end{figure}

The second physically relevant limit are the horizon-crossing values for a general trajectory starting at~$r_0$. These are given by:
\begin{equation}
\mathscr{E}_{\rm hor} = \frac{1}{92 \pi M^2} \left[ \frac{1}{32} \left( 127 + \frac{1}{1-r_0/(2M)} \right) - \frac{6M}{r_0} \right],
\label{energy_unruh_falling_hor}
\end{equation}
\begin{equation}
\mathscr{F}_{\rm hor} = \frac{1}{3072 \pi M^2} \left(97 + \frac{1}{r_0/(2M)-1} - \frac{192 M}{r_0} \right).
\label{flux_unruh_falling_hor}
\end{equation}
We plot these quantities in Fig.~\ref{fig_unruh_hor}. We first notice that they are always finite except in the limit~$r_0 \to 2M$, where~$\mathscr{E}_{\rm hor} \to -\infty$ and~$\mathscr{F}_{\rm hor} \to \infty$. Thus, when released close to the horizon, the observer perceives a huge amount of negative energy entering the black hole. Our interpretation of this behavior is tightly related to the argumentation of what constitutes an Unruh and what constitutes a Hawking effect presented by these authors in~\cite{Barbado2016}, and summarized in~Sec.~\ref{H-vs-U}. The situation just described is one in which separately diverging Hawking and Unruh effects in the outgoing sector interfere negatively leading to a net cancellation. Thus, the outgoing sector does not contribute to the PeRSET. However, the situation in the ingoing sector is absolutely different. In this sector there is no Hawking effect but there exists an enormous Unruh effect due to the diverging acceleration with respect to the asymptotic region. This ingoing sector is the one leading to the divergences of the PeRSET at horizon crossing.

In the limit~$r_0 \to \infty$, the quantities tend to the horizon-crossing values already found for~$\mathscr{E}_{\rm asymp}$ and~$\mathscr{F}_{\rm asymp}$. The energy density is positive except when~$r_0 < \frac{8M}{127} (28 + \sqrt{22}) \simeq 1.03 (2M)$. The flux is always positive and has a minimum at~$r_0= \frac{8M}{95} (24 + \sqrt{6}) \simeq 1.11 (2M)$ of value~$\mathscr{F}_{\rm hor}^{\rm min} = \sqrt{3/2}/(192 \pi M^2)$.

We see now that, contrarily to what happened in the Boulware vacuum, the divergence of the horizon-crossing values is not present, except when the releasing point is arbitrarily close to the horizon itself. These finite results agree with the finite results for the effective temperature functions that we found for the same limit in~\cite{Barbado:2011dx}.

\begin{figure}[h]
\centering
\includegraphics[width=8cm]{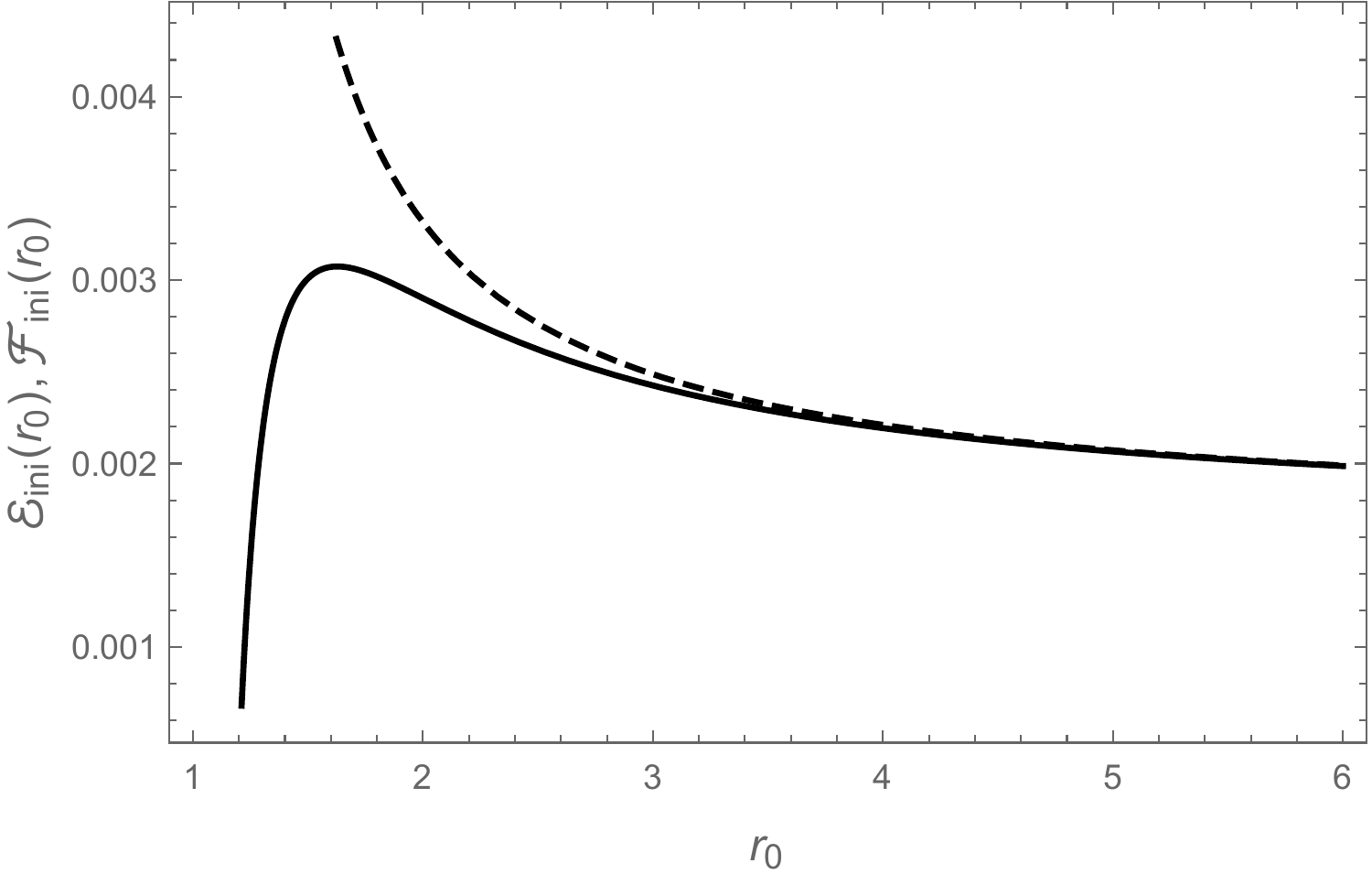}
\caption{Perceived energy density $\mathscr{E}_{\rm ini}$ (solid line) and flux~$\mathscr{F}_{\rm ini}$ (dashed line) as a function of the releasing radius~$r_0$ for free-falling observers in the Unruh vacuum state. We use $2M=1$~units.}
\label{fig_unruh_ini}
\end{figure}

The last physically relevant limit of interest is the starting point itself,~$r = r_0$; that is, the perception of an observer who has just been released into free-fall and thus has zero velocity at a radius~$r_0$. The quantities in this case are:
\begin{equation}
\mathscr{E}_{\rm ini} = \frac{M}{96 \pi r_0^3} \frac{\left[r_0 / (2M)\right]^4 - 2}{r_0/(2M) - 1},
\label{energy_unruh_falling_ini}
\end{equation}
\begin{equation}
\mathscr{F}_{\rm ini} = \frac{1}{768 \pi M^2 ( 1 - 2M/r_0 )}.
\label{flux_unruh_falling_ini}
\end{equation}
We plot them in Fig.~\ref{fig_unruh_ini}. The diverging behavior of these quantities when~$r_0$ approaches the horizon is analogous to that of the horizon-crossing values, and for the very same reason. The energy density now has a maximum at~$r \simeq 1.63 (2M)$ of value~$\mathscr{E}_{\rm hor}^{\rm max} \simeq 0.003/(2M)^2$, before reaching zero and changing sign at~$r_0 = 2^{5/4} M \simeq 1.19 (2M)$.

\subsubsection{Quantum frictionless trajectories}

Quantum frictionless trajectories, introduced in~\cite{Barbado:2015zga}, are those lacking Unruh effect in either the outgoing or the ingoing radiation sector, so that the perception in the corresponding sector is only due to external sources (adequately Doppler and gravitationally shifted). As we first argued in~\cite{Barbado:2015zga} (see also Sec.~\ref{H-vs-U} above and~\cite{Barbado2016}), the Unruh effect should be associated with the acceleration of the observer with respect to the asymptotic region and not with respect to the local free-fall frame. This effect is then encoded in the relative clock rate between the observer and the asymptotic region. Since the natural null coordinates in the asymptotic region are the Eddington-Finkelstein coordinates, the quantum frictionless trajectories for the outgoing~$(-)$ and ingoing~$(+)$ radiation sector are defined by imposing that their relative clock rate is kept constant; that is either or both:
\begin{equation}
\dot{t} \pm \dot{r}^* = Q ( = {\rm const} ),
\label{qft_definition}
\end{equation}
where the dot denotes derivation with respect to proper time.

A nice property of the quantum frictionless trajectories for the outgoing sector is that, in the Boulware vacuum, they satisfy~$\kappa_U = 0$ , and thus also~$\mathscr{E} = -\mathscr{F}$ [for the ingoing sector: $\kappa_V = 0$ and $\mathscr{E} = \mathscr{F}$]. That is, any perception along these trajectories is pure ingoing (outgoing) flux due to the Unruh effect left in the ``other'' sector. Note, however, that this is a necessary condition but not a sufficient condition: The imposition of~$\mathscr{E} = -\mathscr{F}$ implies that~$\dot{\kappa}_U = -\kappa_U^2/2$, which only asymptotically tends to~$\kappa_U = 0$ (and likewise for the ingoing sector).

The family of trajectories of this kind is fully described in~\cite{Barbado:2015zga}---for example, the static trajectories are a particular case, and in fact the only ones which lack Unruh effect in both sectors---. Here, among the set of frictionless trajectories we will only consider ingoing trajectories from the asymptotic region that lack Unruh effect in the outgoing radiation sector. These are given by the expression
\begin{equation}
	r(\tau) = \rf \left[ 1 + W_0 \left(\rme^{- g \tau} \right) \right],
\label{trajectory}
\end{equation}
where $W_0(z)$ is the branch of the Lambert~$W$ function with~$W_0(z) \in \mathds{R}$ and \mbox{$W_0 (z) \geq -1$} for~$z \in [-1/\rme, \infty)$ (see~\cite{Barbado:2015zga}). They reach an asymptotic radius~$\rf$ with asymptotic proper acceleration~$g := M/(\rf^2 \sqrt{1-2M/\rf})$. The interest of these trajectories lies in the fact that they are candidates for a self-consistent buoyancy scenario of a test object falling towards a black hole, due to Hawking radiation pressure~\cite{Barbado:2015zga}.

\begin{figure}[h]
\centering
\includegraphics[width=8cm]{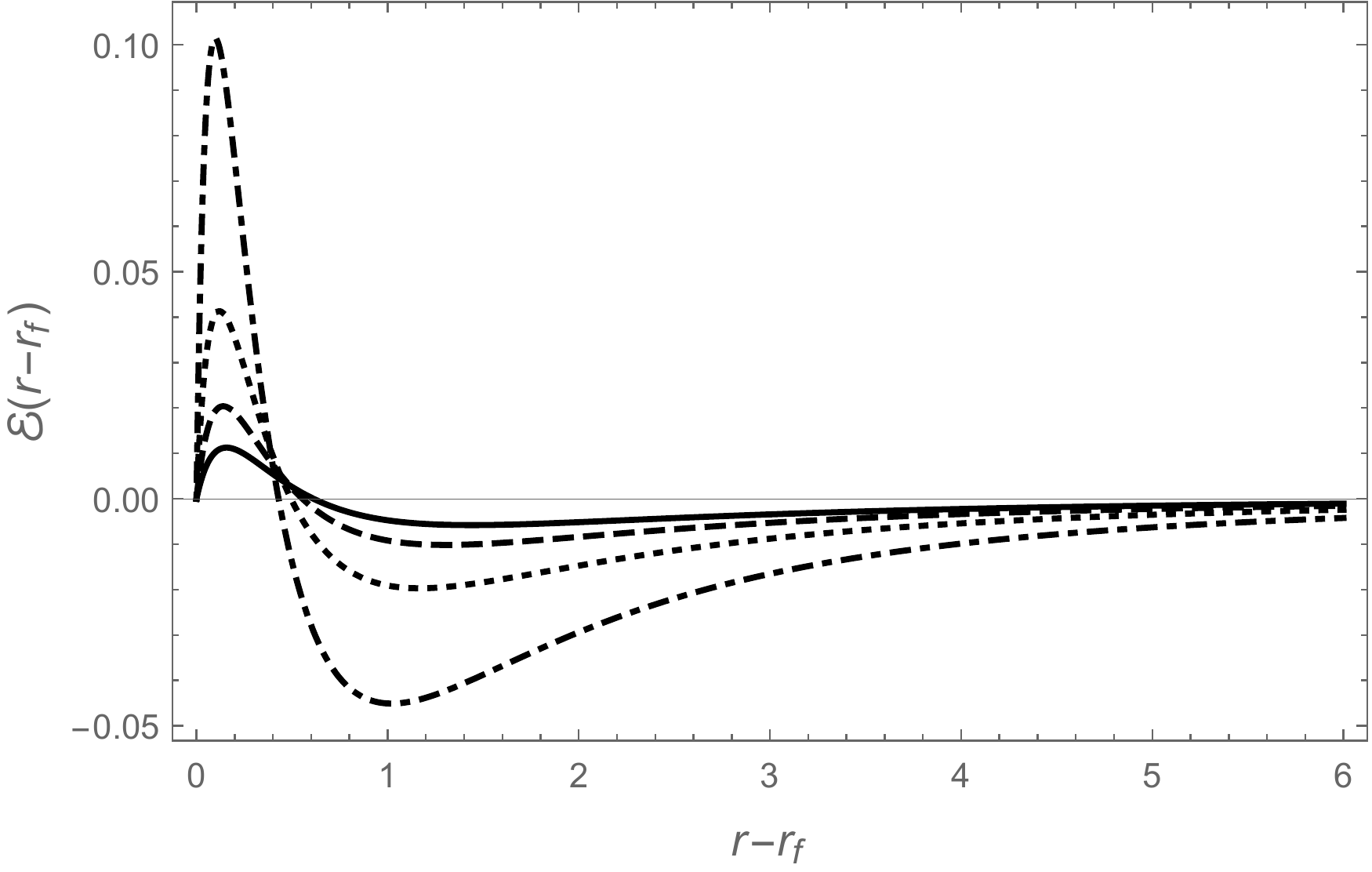}
\caption{Perceived energy density $\mathscr{E}=-\mathscr{F}$ as a function of~$r-\rf$ (for convenience in the plotting) for frictionless trajectories with asymptotic radius~$\rf = $ ($3M$, $2.8M$, $2.6M$, $2.4M$) (solid, dashed, dotted and dash-dot lines, respectively). We use $2M=1$~units.}
\label{fig_frictionless}
\end{figure}

In the Boulware vacuum these trajectories by definition only perceive some Unruh effect in the ingoing radiation sector. Their perceived energy density and flux are given by
\begin{align}
\mathscr{E} = -\mathscr{F} =\ & \frac{M^2 \left(1-\frac{r}{\rf} \right)}{12 \pi r^4 \left(1- \frac{2M}{\rf}\right) \left(1- \frac{2M}{r}\right)^2} \nonumber \\
& \times \Bigg[ \frac{5 M^2}{r^2} - 3 \left(\frac{M}{\rf} + 1 \right) \frac{M}{r} + \frac{2M}{\rf} \Bigg].
\label{qft_boulware}
\end{align}
We plot this quantity for different frictionless trajectories in~Fig.~\ref{fig_frictionless}. Along the ingoing trajectory towards the asymptotic radius $r_f$, the energy density $\mathscr{E}$ first takes negative values, but changes to positive values afterwards, before decreasing back to zero when reaching~$\rf$. The exact location of the positive and negative peak of the perceived energy depends in a complicated way on $r_f$, while their magnitude strongly increases for trajectories closely approaching the horizon (i.e., as $r_f \to 2M$).

When considering the Unruh vacuum instead of the Boulware vacuum, the perceived energy and flux of these frictionless trajectories contains one additional term which corresponds to the constant value of $\kappa_{U,{\rm Haw}}$ in the outgoing sector.

\section{Summary and discussion}\label{sec:summary}

In quantum field theory in curved backgrounds the behavior of a detector coupled to a field depends on the state of the quantum field, but also on the specific trajectory followed by the detector. The techniques used to analyze the \emph{objective} characteristics of the quantum state and the \emph{subjective} appearance of the state to different detectors are somewhat different.
In the first case one uses mainly a tensor, the renormalized stress energy tensor (RSET), while in the second one uses non-tensorial quantities such as Bogoliubov transformations or, equivalently, effective temperature functions.

In this work we have defined a novel renormalized stress energy tensor: the Perception RSET (PeRSET). This object is able to encode in a tensorial manner the relevant information associated with the perception of a specific quantum state by specific observers (or detectors). Then, a proper understanding of a physical situation, with its objective and subjective parts, can be made by analyzing the structure of two parallel tensorial quantities: the RSET, $\langle T_{\mu\nu} \rangle$ and the PeRSET, ${\cal T}_{\mu\nu}$. They contain complementary information.

The PeRSET is defined by subtracting the RSET evaluated in two different vacua: the vacuum state of the field and a vacuum state naturally associated with the specific observer. We show that the PeRSET can be completely expressed in terms of the effective temperature functions (or peeling functions) that have been used in previous perception analyses.  

We have illustrated the physics that can be extracted from the PeRSET by working out several examples of special interest. The obtained results reinforce a novel interpretation advocated by the same authors in~\cite{Barbado2016}, of how to separate the perceived radiation into an Unruh component and a Hawking component. In~\cite{Barbado2016}, we emphasize that crossing the horizon of a black hole is not a simple task when one tries to do it slowly. Here, this can already be understood by looking at the perception of observers at a fixed radial position close to the horizon and those free-falling in the vicinities of the horizon (or, equivalently, those released to free fall towards the horizon from an initial position already close to the horizon). In both situations the perceived energy densities and fluxes are extremely large, and diverge when the fixed radial position or release point approaches the horizon. Thus, independently of their acceleration, detectors moving towards the horizon at small velocities will experience strong vacuum effects, and the horizon will appear as something very different from being a vacuum region.   

The PeRSET defined in this paper should be useful to understand the different analyses on perception scattered throughout the literature in a more ordered and systematic manner. At the same time it will provide additional insight into a topic that is well established but still subject to qualifications and surprises.

\section*{Acknowledgments}

The authors wish to thank an anonymous referee for his/her suggestion, 
which helped to improve this article.
Financial support was provided by the Spanish MINECO through the projects 
FIS2011-30145-C03-01, FIS2011-30145-C03-02, FIS2014-54800-C2-1, FIS2014-54800-C2-2 (with FEDER contribution), and by 
the Junta de Andaluc\'{\i}a through the project FQM219.

\end{document}